\documentstyle[12pt,a4]{article}
\begin{document}
\newcommand{\eq}{\begin{eqnarray}}
\newcommand{\en}{\end{eqnarray}}
\newcommand{\Ms}{M^\star}
\newcommand{\Ps}{P^\star}
\newcommand{\Es}{E^\star}
\newcommand{\mc}{m_\pi^2}
\newcommand{\mo}{m_{\pi^0}^2}
\newcommand{\dpi}{\Delta_\pi}
\begin{center}
{\bf THEORY OF THE $\pi^+\pi^-$ ATOM}
\vskip 5mm

M.A. Ivanov$^{1}$, V.E. Lyubovitskij$^{1,2}$, E.Z. Lipartia$^{3,4}$
and A.G. Rusetsky$^{1,4}$

\vskip 5mm

{\small
(1) {\it
Bogoliubov Laboratory of Theoretical Physics, Joint Institute
for Nuclear Research, 141980 Dubna, Russia
}
\\
(2) {\it
Department of Physics, Tomsk State University,
634050 Tomsk, Russia
}
\\
(3) {\it
Laboratory for Computational Technique and Automation,
Joint Institute for Nuclear Research, 141980 Dubna, Russia
}
\\
(4) {\it
IHEP, Tbilisi State University, 380086 Tbilisi, Georgia
}}
\end{center}

\vskip 3mm
\begin{center}
\begin{minipage}{150mm}
\centerline{\bf Abstract}
The relativistic perturbative approach based on the Bethe-Salpeter (BS)
equation is developed for the study of the
characteristics of the hadronic $\pi^+\pi^-$ atom.
The general expression for
the atom lifetime is derived. Lowest-order corrections to the relativistic
Deser-type formula for the atom lifetime are evaluated within the
Chiral Perturbation Theory.
The lifetime of the $\pi^+\pi^-$ atom in the two-loop order of the
Chiral Perturbation Theory is predicted to be
$\tau_1=(3.03\pm 0.10)\times 10^{-15}~s$.
\end{minipage}
\end{center}

\vskip 7mm

The study of the pion-pion scattering process enables one to gain a deeper
insight in the nature of strong interactions. According to common belief,
low-energy pion dynamics is described in the Chiral Perturbation Theory
(ChPT)~\cite{ChPT} which predicts the $S$-wave $\pi\pi$ scattering lengths
$a_0^0=0.217m_\pi^{-1}$ and $a_0^2=-0.041m_\pi^{-1}$
with an $5\%$ accuracy~\cite{Bijnens}.
The calculations within the Generalized ChPT, involving more parameters,
lead to a most likely values of $\pi\pi$ scattering lengths~\cite{Stern}
which significantly differ from the predictions of the standard ChPT
(e.g. $a_0^0=0.27m_\pi^{-1}$). Despite this difference, both results for the
scattering length $a_0^0$ are compatible with the experimental value
$a_0^0=(0.26\pm 0.05)m_\pi^{-1}$~\cite{Pocanic} due to a large experimental
error. Consequently, a precise measurement of $\pi\pi$ scattering lengths will
be an excellent test of the ChPT.

The existing experimental information on the $S$-wave $\pi\pi$
scattering lengths is extracted from the data on the process
$\pi N\rightarrow\pi\pi N$ and on $K_{e4}$ decays~\cite{Pocanic}.
The DIRAC experiment at CERN is aimed at the
precise measurement of the lifetime of the $\pi^+\pi^-$
atom~\cite{DIRAC}. This will allow to determine the quantity $a_0^0-a_0^2$
with a high accuracy and thus will provide a decisive probe of the
predictions of ChPT. The experiments on the measurement of the hadronic
atom observables are held also at PSI, KEK, Frascati, Uppsala.

The experimental study of $\pi^+\pi^-$ atom characteristics provides a direct
information about the strong $\pi\pi$ amplitudes near threshold since the
average momenta ($\sim \alpha m_\pi$) of constituents in the atom
is much less than either of two typical scales characterizing strong
interactions in chiral theories ($\sim m_\pi$ and $4\pi F_\pi\sim 1~GeV$).
The huge difference between the bound-state and strong scales results in
a clear-cut factorization of strong and electromagnetic interactions in
the observables of the pionium: the Coulomb interaction between pions
is mainly responsible for the atom formation, and strong interactions
lead to its decay. In the lowest-order formula for the ground-state atom
lifetime~\cite{Deser} these two distinct phenomena enter separately through
the value of the nonrelativistic Coulomb ground-state wave function at
the origin $\Psi_1(0)$ and the difference of the $S$-wave scattering lengths
$a_0^0-a_0^2$, respectively.

\eq\label{DESER}
\tau_1^{-1}=\frac{16\pi}{9}\,
\biggl(\frac{2\,\Delta m_\pi}{m_\pi}\biggr)^{{1}/{2}}
(a_0^0-a_0^2)^2\,\,|\Psi_1(0)|^2
\en
where, using the values of the scattering lengths from the two-loop
calculations in standard ChPT $a_0^0-a_0^2=0.258m_\pi^{-1}$, we obtain
$\tau_1=3.20\times 10^{-15}~s$.

In the view of the forthcoming DIRAC high-precision experiment which
is aimed at the extraction of the difference $a_0^0-a_0^2$ from the
pionium lifetime measurement, it is important to estimate the precision
of Eq.~(\ref{DESER}) itself and calculate all possible corrections
to it which were neglected in the original derivation.
In particular, one can readily observe from Eq.~(\ref{DESER}) that the
{\it ad hoc} change of the "reference mass" from the charged to the neutral
pion mass in the definition of strong scattering lengths leads to a
$\sim 12\%$ variation in the lifetime. The leading electromagnetic corrections
of order $O(\alpha)$ will also be important in the analysis of the
high-quality data from DIRAC experiment.

The earlier efforts of the evaluation of the corrections to the pionium
lifetime in the field-theoretical framework~\cite{History} were focused
on the evaluation of the quantities entering into Eq.~(\ref{DESER}).
These investigations resulted in the understanding of the fact that
the consistent approach to the calculation of bound-state characteristics
can not be confined solely to the corrections to the scattering lengths
and atomic wave function. Instead, to this end a systematic field-theoretical
framework is needed to consistently take into account the pion strong
dynamics described by the effective ChPT Lagrangian in order to provide
a reliable and unambiguous test of the predictions of ChPT.
In the several recent papers
an attempt was made to carry out a comprehensive field theoretical analysis
of the pionium problem. Namely, in~\cite{Sazdjian} on the basis of the
3-dimensional constraint theory bound-state equations the mass shift,
radiative and second-order strong corrections have been evaluated.
In our previous papers~\cite{Atom} a systematic perturbative approach based
on the BS equation is constructed, and a complete set of the
lowest-order corrections to the pionium lifetime is evaluated. The vacuum
polarization effect on the pionium observables is evaluated in~\cite{Labelle}
within the nonrelativistic QED.

In a nonrelativistic potential approach to the problem~\cite{Rasche}
the expression for the lifetime, including the corrections, is written
in terms of physical scattering lengths in charged and neutral channels
in the presence of Coulomb interactions. This is a counterpart of the
lowest-order Deser-type factorization: the bound-state observables
are expressed in terms of the quantities characterizing the scattering
process, and these quantities differ from the purely strong ones by
the mass shift and electromagnetic corrections.
In order to evaluate these corrections, one needs the $\pi\pi$
potential to be given explicitly. Assuming local and energy-independent
strong potentials in the isospin symmetry limit, one reproduces the
$\pi\pi$ scattering phase shifts calculated from ChPT. At the next step
the Coulomb potential is added and the mass shift effect is
included via assigning the physical masses to pions, the strong potential
remaining unchanged.

Below we give the basic ideas and assumptions of our approach to the
pionium observables, based on the BS equation.

The pions within our approach are described by the elementary fields
in the Lagrangian. This allows one to write down the bound-state
BS equation for the metastable $\pi^+\pi^-$ atom. The necessary link
to the strong amplitudes from ChPT where the pions are described by
the pseudoscalar quark densities, proceeds via the Deser-type factorization
in analogy with the potential theory. Namely, we express all corrections to
the Deser formula in terms of the on-mass-shell pion amplitudes which are
further identified with the amplitudes from ChPT.

In our calculations we always adopt the so-called "local" approximation
which consists is suppressing the relative momentum dependence of the
strong amplitudes. The origin of this approximation can be traced back
to the huge difference between the bound-state and strong momentum scales.

The pionium in our approach is described by the exact bound-state
BS equation with the full kernel $V(P)$ which, apart from all two-particle
irreducible diagrams includes the self-energy corrections in two
outgoing charged pion legs~\cite{Atom}. The square of the
c.m. momentum $P$ takes the value $P^2=\bar M^2=M^2-iM\Gamma$ where $M$
denotes the "mass" of an atom, and $\Gamma$ stands for the decay width.

In order to perform the perturbative expansion of the bound-state
observables we single out from the full kernel $V$ the instantaneous
Coulomb part $V_C$ which
is responsible for the formation of the bound state composed of
$\pi^+$ and $\pi^-$. The "remainder" of the potential which is treated
perturbatively is denoted by $V'=V-V_C$.

The "unperturbed" part of the full kernel $V_C$ is chosen in the
form guaranteeing the exact solvability of the corresponding BS equation.
The solution of this equation yields the relativistic Coulomb wave function
$\psi_C$ with the corresponding eigenvalue
${\Ps}^2={\Ms}^2=m_\pi^2(4-\alpha^2)$ where $m_\pi$ is the charged pion mass.
Then, the solution of the full BS equation
is expressed in terms of the unperturbed solution~\cite{Atom}.
Substituting this solution into the complete BS equation
we arrive at the relation
$$
<\psi_C|\bigl[ 1+(G_0^{-1}(P)-G_0^{-1}(\Ps)-V'(P)) G_R Q\bigr]^{-1}
(G_0^{-1}(P)-G_0^{-1}(\Ps)-V'(P))|\psi_C>=0
$$
which serves as the basic equation for performing the perturbative expansion
of the bound-state observables.
Here $G_0$ denotes the free Green's function of the $\pi^+\pi^-$ pair,
$G_R$ stands for the pole-subtracted part of the exact relativistic
Coulomb Green's function and $Q$ is the projection operator on the
subspace orthogonal to the unperturbed ground-state BS wave
function~\cite{Atom}.
The explicit form of the operator $G_RQ$
is given and the perturbation kernel $V'$ is known up to any given
order of loop expansion. The only unknown quantities entering
into this expression are the real and imaginary parts of $P^2$.
Consequently, expanding this expression in powers of $V'$, we arrive at the
recursive relations which define the mass and width of the bound state
in any given order.

In order to provide the necessary Deser-type
factorization in the obtained results, one needs a further classification
of diagrams entering into the definition of $V'$. We consider the
following subsets of diagrams:

\noindent
1. A purely strong part, which is isotopically invariant.

\noindent
2. The part which is responsible for the $m_{\pi^\pm}-m_{\pi^0}$
electromagnetic mass difference

\noindent
3. Remaining electromagnetic effects, including the exchanges of
virtual photons.

The sum of the potentials corresponding to the parts 1 and 2 is denoted by
$V_{12}=V_1+V_2$. The $T$-matrix
corresponding to summation of the potential $V_{12}$ in all orders
is given by $T_{12}(P)=V_{12}(P)+V_{12}(P)G_0(P)T_{12}(P)$. The rest
of the potential is referred to as $V_3=V'-V_{12}$ and is treated
perturbatively. Further, the quantity $G_RQ$ in our basic relation can
be written as $G_RQ=G_0(\Ps)+\delta G$ where $\delta G$ corresponds to
the exchange of the ladder of Coulomb photons and is also considered
as a perturbation. Then in the first order in $V_3$ and $\delta G$ the basic
relation takes the form
\eq\label{LARGE}
&&0=-2i\Ms\delta M-<\psi_C|T_{12}|\psi_C>+\nonumber\\[2mm]
&&+\,\delta M<\psi_C|(G_0^{-1})'G_0T_{12}|\psi_C>
+\frac{1}{2}(\delta M)^2<\psi_C|(G_0^{-1})''+2T_{12}G_0(G_0^{-1})''|\psi_C>+
\nonumber\\[2mm]
&&+<\psi_C|(\delta M(G_0^{-1})'-T_{12})\delta GT_{12}|\psi_C>
-<\psi_C|(1+T_{12}G_0)V_3(1+G_0T_{12})|\psi_C>\quad\quad
\en
where $G_0=G_0(\Ms)$ and the prime stands for the differentiation with
respect to $\Ms$.

In the "local" approximation $T_{12}$ does not depend on the relative
momenta and thus can be taken out of the matrix elements.
Neglecting first the
corrections (only the first line of Eq.~(\ref{LARGE}) contributes),
we arrive at the relativistic counterpart of Eq.~(\ref{DESER})
\eq\label{rel-deser}
\biggl\{
\begin{array}{c}
\Delta E^{(1)}\cr -1/2\,\,\Gamma^{(1)}
\end{array}
\biggr\}
=\frac{16\pi i}{2\Ms m_\pi}|\Psi_1(0)|^2
\biggl\{
\begin{array}{c}
{\mbox{\rm Re}}\cr {\mbox{\rm Im}}
\end{array}
\biggr\}\,\,
{\cal T}_{\pi^+\pi^-\rightarrow\pi^+\pi^-}(4m_\pi^2;\vec 0,\vec q_0,)
\en
where ${\cal T}(s;\vec p,\vec q\,)=16\pi T_{12}$ denotes the $S$-wave
$\pi\pi$ strong scattering amplitude which includes the effect of
$m_{\pi^\pm}-m_{\pi^0}$ mass difference. $\vec q_0$ is the relative momentum
of the $\pi^0\pi^0$ pair at the threshold $s=4m_\pi^2$, with the magnitude
given by the relation $m_\pi^2=m_{\pi^0}^2+\vec q_0^{~2}$.

In order to calculate the corrections to the relativistic Deser formula,
one has to evaluate the integrals entering into the Eq.~(\ref{LARGE}).

\noindent {\bf 1.}
The correction due to the shift of the bound-state pole by strong
interactions is determined by two terms in the second line of Eq.~(\ref{LARGE}).
Calculating these integrals explicitly and using Eqs.~(\ref{rel-deser}) we
obtain
\eq
\delta_S=-9\Delta E^{(1)}/(8E_1)=
-5.47\times 10^{-3}\,\,m_\pi(2a_0^0+a_0^2)
\en
where $E_1$ denotes the unperturbed ground-state binding energy.

\noindent {\bf 2.}
The correction due to the relativistic modification of the Coulomb wave
function is given by the value of the wave function $\psi_C$ at the origin
which for the particular choice of the kernel used in the calculations
is related to its nonrelativistic counterpart by
\eq
\psi_C(0)=\Psi_1(0)\,\, (1-0.381\alpha+\cdots)
\en
and the correction in the decay width which is proportional
to $|\psi_C(0)|^2$
is twice as large.

\noindent {\bf 3.}
The correction due to the exchange of Coulomb photon ladders is given
by the first term in the third line of Eq.~(\ref{LARGE}) containing
$\delta G$. Using the known expression for the relativistic Coulomb
Green's function~\cite{Atom} and evaluating this integral explicitly,
we obtain
\eq
\delta_C=(1/2+2.694-{\rm ln}\alpha)\,\Delta E^{(1)}/E_1=
3.95\times 10^{-2}\,\,m_\pi(2a_0^0+a_0^2)
\en

\noindent {\bf 4.}
In order to obtain the mass shift and radiative corrections, in
the perturbation kernel $V_3$ one has to take into account
the residual photon exchange diagram and the self-energy corrections
in the outgoing charged pion legs. The net result of this effect consists
in the cancellation of the correction due to the relativistic
modification of the Coulomb wave function
and in replacing the quantity ${\cal T}_{\pi^+\pi^-\rightarrow\pi^0\pi^0}$
which appears in the imaginary part of the $\pi\pi$ scattering
amplitude in Eq.~(\ref{rel-deser}), by the Coulomb pole removed
full amplitude for the process $\pi^+\pi^-\rightarrow\pi^0\pi^0$.
Using the expression of this amplitude~\cite{Knecht}, and isolating the purely
strong isotopically symmetric amplitude with a common mass equal to
the charged pion mass, we arrive at the following expressions for the
mass shift and radiative corrections
\eq
\delta_M=\frac{2\Delta m_\pi^2}{3m_\pi^2}
\biggl(1+\frac{m_\pi^2}{96\pi^2F_\pi^2}
\biggl(48+\frac{16}{3}\bar l_1-\frac{16}{3}\bar l_2+7\bar l_3-36\bar l_4
\biggr)\biggr)
\en
\eq
\delta_{em}=\alpha/(12\pi)\,
(-30+3{\cal K}_1^{\pm 0}-{\cal K}_2^{\pm 0})
\en
where ${\cal K}_i^{\pm 0}$ and $\bar l_i$ denote the low-energy constants
of ChPT~\cite{ChPT,Knecht}.

\noindent {\bf 5.}
Taking into account the diagram in the kernel $V_3$ corresponding to
the vacuum polarization effect and evaluating numerically
the resulting integral, we arrive at the following result
\eq
\delta_{vac}=3\alpha^2\, m_\pi/(16\,m_e)\times 0.6865
\en
which completely agrees with the zero-Coulomb piece of the result
given in Ref. \cite{Labelle}.

\noindent {\bf 6.}
The correction to the pionium decay width due to the finite size
effect is caused by the modification of the instantaneous Coulomb
interaction by the pion loop in the two pion-photon vertex in the
kernel $V_3$. For simplicity,
we approximate this vertex by a monopole parameterization.
After calculating the corresponding matrix element explicitly, we obtain
\eq
\delta_F=2\alpha\mc<r^2>_V^\pi/(3\pi)\,\,
{\rm ln}\,(\mc<r^2>_V^\pi/24)
\en
where $<r^2>_V^\pi$ denotes the square charge radius of the pion.

The relativistic Deser formula with all above corrections takes the form
$$
\tau_1^{-1}=\frac{16\pi}{9}\,
\biggl(\frac{2\,\Delta m_\pi}{m_\pi}\biggr)^{{1}/{2}}
\biggl(1-\frac{\Delta m_\pi}{2m_\pi}\biggr)^{1/2}
(a_0^0-a_0^2)^2\,\,|\Psi_1(0)|^2\,\,
(1+\delta_S+\delta_C+\delta_M+\delta_{em}+\delta_{vac}+\delta_F)
$$
where $a_0^0$ and $a_0^2$ denote the $\pi\pi$ scattering lengths
in the isospin-symmetric case, with the charged pion mass taken
to be the common mass of the pion isotriplet.

In our calculations for the constants $\bar l_i$ we take the
numerical values from Ref. \cite{ChPT}
$\bar l_1=-2.3\pm 2.7$, $\bar l_2=6.0\pm 1.3$, $\bar l_3=2.9\pm 2.4$,
$\bar l_4=4.3\pm 0.9$. Also, we use the values
$\frac{e^2F_\pi^2}{\mc}\,{\cal K}_1^{\pm 0}=1.8\pm 0.9$,
$\frac{e^2F_\pi^2}{\mc}\,{\cal K}_2^{\pm 0}=0.5\pm 2.2$
from Ref.~\cite{Knecht}.
Also, we adopt the following value of the e.m. charge
radius of pion $<r^2>_V^\pi=0.439~Fm$ \cite{ChPT}.
With the use of the above values of the parameters the total correction
to the pionium decay width turns out to be $\delta_{tot}=(6.1\pm 3.1)\%$
and
$$
\tau_1=(3.03\pm 0.10)\times 10^{-15}~s
$$
The largest correction in the decay width is caused by the
total effect of the mass shift and electromagnetic radiative
corrections.
The recent field-theoretical calculations carried in the framework
of the 3D relativistic equations have come to a general agreement
with our results for all individual effects.
However, the sign of the mass shift effect obtained in the nonrelativistic
scattering theory approach \cite{Rasche} turns out to be opposite
as compared to our result, and is of the same order of magnitude.
Owing to the derivative character of pion couplings in the chiral Lagrangian,
a possible reason for this discrepancy might be an explicit energy
independence of strong potentials used in these calculations.

To summarize, we have evaluated a complete set of the lowest-order corrections to
the pionium decay width in the "local" approximation for the strong $\pi\pi$
amplitude. For a full understanding of the problem,
however, the reason for the difference in sign in the mass shift effect
in the field-theoretical and potential theories should be investigated in
detail. Also, it is very important to have a reliable quantitative estimate of the
accuracy of "local" approximation including a proper treatment of arising
new UV divergences. Note that recent attempts of taking into account
the momentum dependence of strong amplitudes have so far led to a controversial
results for the correction to the decay width $\sim 8\%$~\cite{Kong},
$\sim 2\%$~\cite{Bunatian} (large values of the pionium lifetime obtained
in the latter paper are merely due to the use of the tree-level scattering
lengths in the Deser formula).


{\it Acknowledgments}.
We thank V. Antonelli, A. Gall, A. Gashi, J. Gasser,
E.A. Kuraev, H. Leutwyler, P. Minkowski, L.L. Nemenov, E. Pallante, H. Sazdjian
and Z. Silagadze
for useful
discussions, comments and remarks. A.G.R. thanks Bern University for
the hospitality where part of this work was completed.
This work was supported in part by the Russian Foundation for
Basic Research (RFBR) under contract 96-02-17435-a.


\end{document}